\documentclass[debug]{rmaa}

\usepackage{paralist}

\usepackage{psfrag,color}

\usepackage[latin1]{inputenc}

\usepackage{graphicx}

\newcommand{\Msun}{\mbox{M$_{\odot}$}}
\newcommand{\Rsun}{\mbox{R$_{\odot}$}}

\newcommand{\kms}{\mbox{km\,s$^{-1}$}}

\title{Solar-type eclipsing binary KIC\,4832197: Physical properties and intrinsic variability of the components}

\author{
  O. \"Ozdarcan,\altaffilmark{1} 
  H. A. Dal,\altaffilmark{1}
  and E. Yolda\c{s}\altaffilmark{1}}

\altaffiltext{1}{$^1$Department of Astronomy and Space Sciences, 
Faculty of Science, Ege University, 35100 , {\.I}zmir, Turkey.}

\shortauthor{\"Ozdarcan, Dal \& Yolda\c{s}}
\shorttitle{Solar-type eclipsing binary KIC\,4832197}

\fulladdresses{
\item O. \"Ozdarcan, H. A. Dal and E. Yolda\c{s}: Ege University, Faculty of Sciences, 
Department of Astronomy and Space Sciences, 35100 Bornova, \.{I}zmir, Turkey.

}

\listofauthors{O. \"Ozdarcan, H. A. Dal \& E. Yolda\c{s}}
\indexauthor{\"Ozdarcan, O.}
\indexauthor{Dal, H. A.}
\indexauthor{Yolda\c{s}, E.}

\abstract{Comprehensive analysis of optical spectroscopy and 
space photometry of the solar type eclipsing binary system 
KIC\,4832197 is presented. The system is composed of F7V + F9V 
components with masses of $M_{1}=1.16\pm0.12~\Msun$, 
$M_{2}=1.07\pm0.10~\Msun$ and radii of $R_{1}=1.26\pm0.04~\Rsun$, 
$R_{2}=1.03\pm0.03~\Rsun$. Position of the components on 
$Log~T_{eff}-Log~L/L_{\odot}$ plane suggests an age of 
$2.8\pm0.8~Gyr$ for the system. Inspection of out-of-eclipse 
brightness in time reveals wave-like variability pattern, whose 
amplitude and shape quickly change in order of days. Frequency 
analysis of this variability results in two significant peaks in 
amplitude spectrum, which are interpreted as rotational modulation 
of spots on the components. Assuming both spots are on the same 
component, a lower limit for differential rotation coefficient 
is computed as $k=0.12$, which is weaker compared to the solar 
value of $k_{\odot} = 0.189$.}

\resumen{}

\addkeyword{(stars:) binaries: eclipsing}
\addkeyword{stars: fundamental parameters}
\addkeyword{stars: individual (KIC\,4832197)}
\addkeyword{stars: activity}

\begin{document}
\maketitle

\section{Introduction}\label{Sec_intro}

Observed rotational modulation of brightness in a light 
curve of a solar-type star is interpreted as a strong 
photometric evidence of cool star spots on the stellar 
surface, which co-rotate with the surface. These 
spots might emerge in various locations on the surface 
of the star as the manifestation of the magnetic activity, 
which is commonly observed among solar-type stars. Such 
a light curve allows one to trace temporal and spatial 
evolution of spots, as well as determining photometric 
period of the star. Finding photometric period of a 
single star is crucial because that period can be 
considered as the rotation period the star. In the case 
of binary stars, it is possible to determine photometric 
period and orbital period separately. If the components of
the binary system are solar-type stars, then comparison 
between the photometric period and the orbital period
provides hints on surface differential rotation, which 
is one of the key parameter that drives dynamo mechanism 
in stellar interiors.

Until the era of very high precision space photometry, 
photometric studies of the magnetic activity on 
solar-type stars had relied on long-term time-series 
photometry obtained from the dedicated ground-based telescopes 
\citep{APTs_Henry_1995ASPC, APTs_Strassmeier_1997PASP, 
APTs_Rodono_2001AN....322..333R} or all sky surveys 
\citep{ASAS_Pojmanski_1997, ASAS_SN_2017PASP..129j4502K}.
These sources provided photometric data with an uncertainty
of a few percent of magnitude \citep[rarely in milimag level; 
e.g.][]{APTs_Henry_1995ASPC} and enabled to trace short and 
long-term magnetic activity behaviour of solar-type stars in 
terms of mean brightness, light curve amplitude and the 
photometric period.

After entering the era of groundbreaking space photometry, 
astronomers met extremely high precision photometric data 
obtained in a wide wavelength range. Especially, $Kepler$ 
space telescope \citep{Kepler_2010Sci...327..977B, 
Kepler_K2_2014PASP..126..398H} reached photometric 
precision down to a few tens of part-per-million. Such a 
precision not only enabled to detect a planetary transit 
in a light curve, which is primary mission of $Kepler$ 
space telescope, but also sub-milimag amplitude variability 
of stars. In the case of magnetic activity of solar-type 
stars, such precision allowed detection of very small 
amplitude rotational modulation of brightness, which could 
not be distinguished in ground-based photometry due to the 
typical observational scatter of a few per-cent magnitude. 
In addition to its very high precision, $Kepler$ photometry 
spans over four years without any considerable time gap, 
which enables one to trace low amplitude photometric signs 
of magnetic activity of solar-type stars. All these 
properties encouraged researchers for more detailed and 
comprehensive photometric studies on stellar flare and 
differential rotation on large sample of stars 
\citep{Balona_Flare_2015MNRAS.447.2714B, 
reinhold_2015A&A...583A..65R} or on individual targets, 
\citep[particularly eclipsing binaries;][]
{Yoldas_Dal_V1130Cyg_V461Lyr_2021RMxAA..57..335Y, 
Ozdarcan_kic9451096_2018RMxAA..54...37O}. In some cases, 
$Kepler$ photometry reveals effects of two separate 
variability mechanisms, (e.g. pulsation and cool spot 
activity), in an eclipsing binary system 
\citep[see, e.g.,][]{Ozdarcan_kic7385478_2017PASA...34...17A}.

In this study, we present comprehensive analysis of a 
solar-type eclipsing binary system KIC\,4832197. Our 
analysis is based on ground-based medium resolution optical 
spectroscopy and $Kepler$ photometry. KIC\,4832197 takes 
attention with its very shallow eclipse depths in its light 
curve and remarkable out-of-eclipse variability with an 
amplitude that is comparable to the eclipse depths. Neither 
eclipses nor out-of-eclipse variability of the system has 
been discovered until the advent of very high precision 
$Kepler$ photometry. The system was included in the planet 
candidate catalogue due to its shallow eclipse depths 
\citep{planet_catalogue_2016ApJS..224...12C}. However no 
confirmed exoplanet in this system has been reported so far. 
KIC\,4832197 appeared as late A spectral type star according 
to Tycho-2 measurements \citep[$B-V=0\fm202\pm0\fm111$,][]
{Tycho_Hog_et_al_2000A&A...355L..27H}. On the other hand, 
more recent and precise broad-band $UBV$ colours and 
magnitudes of the system are $V=11\fm673\pm0\fm018$, 
$B-V=0\fm496\pm0\fm030$ and $U-B=-0\fm002\pm0\fm032$ 
\citep{ubv_kepler_2012PASP..124..316E}, which indicate
F7 spectral type for the system \citep{Gray_2005}.

Remarkable out-of-eclipse variability may indicate cool spot
activity or pulsations on one or both components. Furthermore, 
shallow eclipses may indicate a possible third light, which may 
reduce eclipse amplitudes in the light curve depending on its 
contribution to the total light of the system. All these properties 
of KIC\,4832197 are promising not only for testing stellar evolution 
models alone, but also for further studies on the variability mechanism 
currently running in one or both components. Analysis of very high
precision $Kepler$ photometry is excellent for such studies.

In this context, we combine ground-based optical spectroscopy and 
space photometry to determine physical properties of the system 
and its evolutionary status. We further analyse brightness 
variability at out-of-eclipse phases, which can shed light into 
the currently running variability mechanism on the components of 
KIC\,4832197. We organize the remaining parts of our study as 
follows. We describe observational data and reduction in the 
next section. Section~\ref{Sec_analysis} includes light time 
variation analysis, spectroscopic and photometric modelling of 
the system, including atmospheric parameter estimation of each 
component, spectroscopic orbit of the system and light curve 
analysis. In the last section, we summarize our findings 
and give discussion via analysis results.

\section{Data}\label{Sec_data}

\subsection{Spectroscopy}\label{Sec_data_spec}

Medium resolution optical spectrum of KIC\,4832197 were 
obtained at T\"UB\.ITAK National Observatory (TNO) with 
1.5 m Russian -- Turkish telescope and Turkish Faint Object 
Spectrograph Camera 
(TFOSC\footnote{\url{https://tug.tubitak.gov.tr/en/teleskoplar/rtt150-telescope-0}}).
An Andor DW436-BV 2048 $\times$ 2048 pixels CCD camera 
with a pixel size of 13.5 $\times$ 13.5 $\mu m^{2}$ was 
used in observations, which allows recording optical 
spectra between 3900 -- 9100 \AA\@ in 11 \'echelle orders. 
This instrumental set-up provided an average resolution 
of $R = \lambda/\Delta\lambda = 2700 \pm 500$ around 
$\lambda=5500$ \AA\@ wavelength region. Observations were
carried out eight nights between July 2014 and April 2017. 
Ten spectra were recorded in total.

Conventional procedures for reducing \'echelle spectrum 
images are followed. These steps are applied under IRAF 
environment\footnote{The Image Reduction and Analysis 
Facility is hosted by the National Optical Astronomy 
Observatories in Tucson, Arizona at URL iraf.noao.edu} 
and starts with removal of bias level and continues by 
division of bias-removed object and calibration lamp images 
by normalized flat-field image. Then, scattered light 
correction and cosmic rays removal steps are applied to 
the bias and flat-field corrected images. Finally, object 
and calibration lamp spectra are extracted from the \'echelle 
orders. Wavelength calibration of extracted object spectra 
is done by using a Fe-Ar calibration lamp spectrum 
recorded at the same observing night. After completing 
standard reduction steps, all object spectra are 
normalized to the unity by applying 4th or 5th order 
cubic spline functions.

\subsection{$Kepler$ photometry}\label{Sec_data_phot}
\begin{figure}
	\includegraphics[scale=0.73]{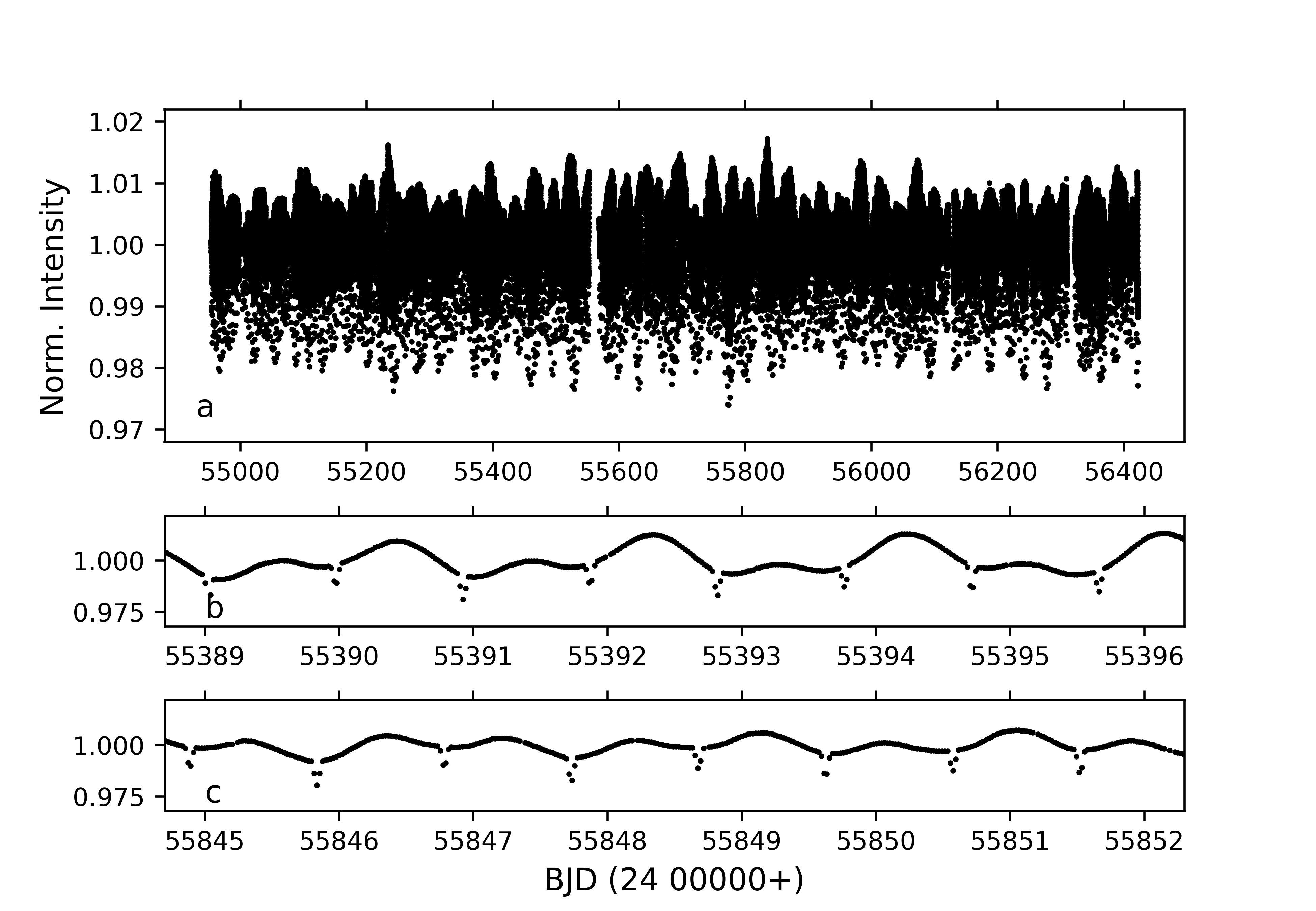}
	\caption{In panel a, long cadence light curve of 
	KIC\,4832197 is shown. In panels b and c, different
	portions of the long cadence light curve are shown, 
	where the shape of the light curve remarkably changes.}
    \label{Fig_all_LC}
\end{figure}

Very high precision space photometry of KIC\,4832197 
was obtained by $Kepler$ spacecraft. The spacecraft
recorded images in 6.02 second exposure time with 0.52
second read-out time in a broad wavelength range 
between 4100 \AA~ and 9100 \AA~
\citep{Kepler_data_characteristics_Gilliland_et_al_2010ApJ...713L.160G}.
Broad wavelength range allows to record more photons in a 
single exposure, which increases precision, at the expense 
of losing colour information. From recorded images, two 
data sets were created depending on two separate integration 
times; 58.9 second (short cadence data) and 29.4 minute 
(long cadence data). These integrations were grouped as 
separate data sets, where each set covered approximately 
three months and called as quarter, except the first quarter, 
which covered ten days of commissioning phase and called 
quarter zero (Q0). During operation of the spacecraft 
over 4 years (between 2009 and 2013) photometric data 
were collected for eighteen quarters in total (from Q0 
to Q17). Continuous long cadence photometry obtained in
each quarter is available for the majority of $Kepler$ 
targets, including KIC\,4832197. Long cadence photometry 
of KIC\,4832197 for each quarter is obtained from Mikulski 
Archive for Space Telescopes (MAST). Prior to analyses,
it is necessary to remove instrumental effects from the
long cadence data. For each quarter, simple aperture 
photometry (SAP) fluxes are considered. These fluxes are 
de-trended as described in \citet{Slawson_et_al_2011} 
and \citet{Prsa_et_al_2011}. De-trended fluxes are then 
normalized to the unity. Analyses and modelling process 
are based on these normalized fluxes. We show the long 
cadence light curve of KIC\,4832197 in 
Figure~\ref{Fig_all_LC}.

\section{Analysis}\label{Sec_analysis}

\subsection{Light time variations}\label{Sec_analysis_light_time}

The first step of the analysis is to determine mid-eclipse 
times of KIC\,4832197 for primary eclipses. In principle, 
mid-eclipse times can be determined straightforwardly by 
applying Kwee-van Woerden method 
\citep{Kwee_Van_Woerden_1956BAN....12..327K} to observations 
close by to mid-eclipse time. An alternative way is to fit a 
function to the observational data around estimated mid-eclipse 
time and determine the extremum point of this function, which 
corresponds to the mid-eclipse time. These methods may 
work flawlessly for a light curve of an ordinary eclipsing 
binary, which shows symmetric eclipse light curves with 
respect to the mid-eclipse time and exhibits flat or slightly 
distorted maxima at out-of-eclipse phases. Such light curves 
indicate the absence of intrinsic variability for any of the 
component. Looking at Figure~\ref{Fig_all_LC}, one may easily 
notice the variable nature of the light curve. Because of the 
relatively short orbital period and integration time of
long cadence data, a few data points can be found around the
expected mid-eclipse time for a given orbital cycle. Combination of
these two effects gives a complex shape to the light curve
and makes the methods mentioned above improper for precise
determination of mid-eclipse times. In such cases, one of 
the most reliable way is to find a best-fitting light curve 
model and then only adjust mid-eclipse time of the primary 
minimum in the model for each cycle. We follow exactly this 
way for precise determination of mid-eclipse times from long 
cadence data. 

Actually, this is an iterated process, which starts by 
preparing a phase-folded light curve with initial light 
elements, i.e. an ephemeris reference time ($T_{0}$) and 
an orbital period ($P$), then continues by finding the
best-fitting light curve model. In the case of KIC\,4832197, 
initial $T_{0}$ and $P$ values are adopted from Kepler 
Eclipsing Binary Catalogue\footnote{http://keplerebs.villanova.edu/} \citep{Prsa_et_al_2011, Slawson_et_al_2011}
given in Equation~\ref{Eq1}:

\begin{equation}
    T_{0} {\rm (BJD)} = 2,454,954.965798 + 1\fd8954655 \ \times \ E .
	\label{Eq1}
\end{equation}

After finding the best-fitting light curve model, 
$T_{0}$ is adjusted for each orbital cycle separately, by 
keeping all other model parameters fixed. In this step,
time based long cadence data are considered, instead of
phase-folded light curve. Thus, mid-eclipse time of each 
orbital cycle can be determined precisely together with 
formal errors. Practical application of this method is 
done by Wilson-Devinney code (see later sections for details). 
After determination of mid-eclipse times, a simple eclipse 
time variation (etv) diagram is constructed and linear 
corrections are determined and applied to both $T_{0}$ 
and $P$. Then the whole process is repeated by using 
corrected $T_{0}$ and $P$. In most cases, two iterations 
are fairly enough to obtain a self consistent $T_{0}$, $P$ 
and light curve model parameters. Converged solution leads 
to the corrected light elements given in Equation~\ref{Eq2}:

\begin{equation}
    T_{0} {\rm (BJD)} = 2,454,954.9661(2) + 1\fd8954650(4) \ \times \ E .
	\label{Eq2}
\end{equation}

\begin{figure}
	\includegraphics[scale=0.73]{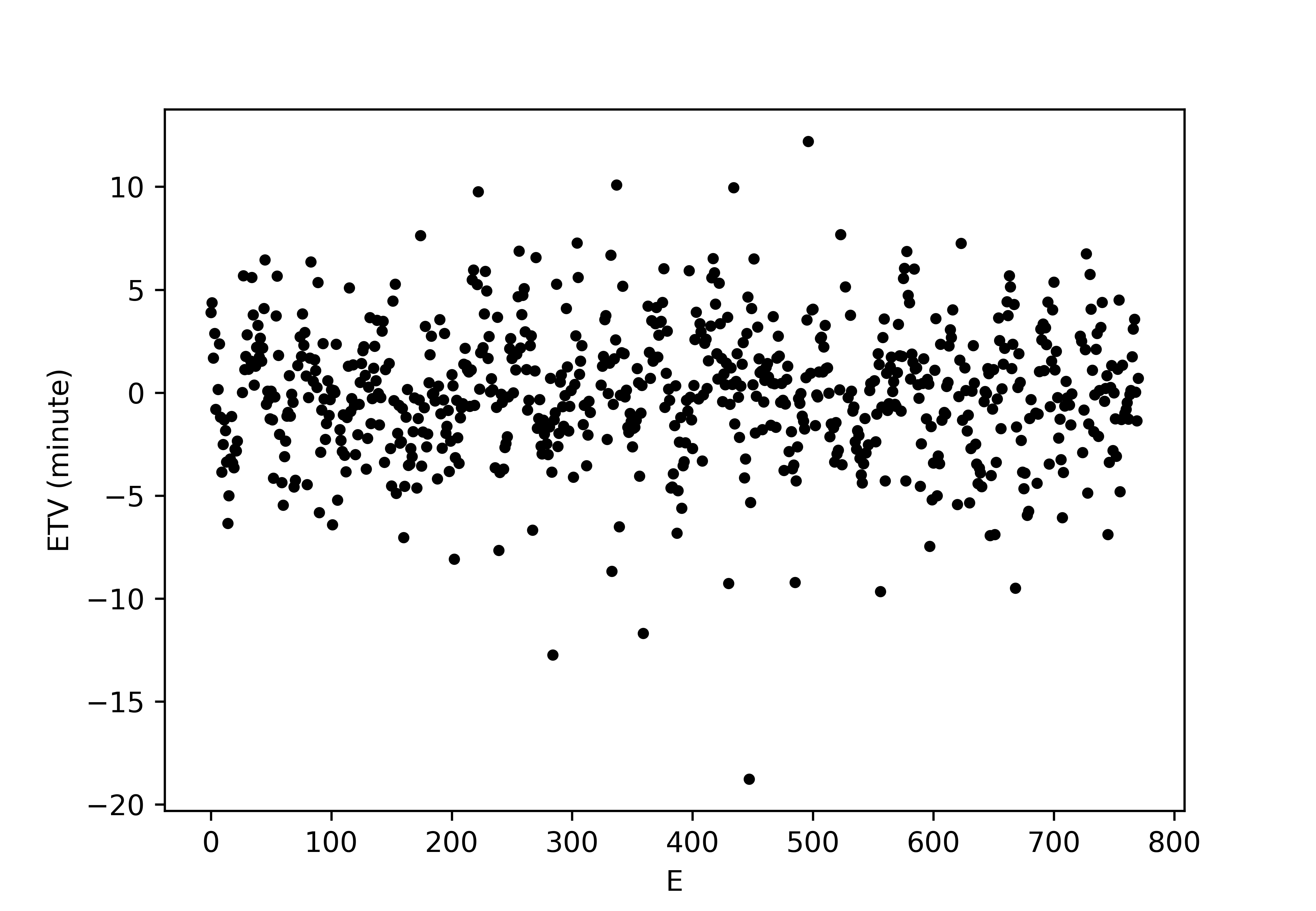}
	\caption{ETV diagram for KIC\,4832197.}
    \label{Fig_etv}
\end{figure}

The numbers in parentheses are shown statistical uncertainties 
for the last digit of corresponding parameter. These 
uncertainties are computed from linear least squares fit. 
$T_{0}$ and $P$ values given in Equation~\ref{Eq2} are adopted 
for further analyses and kept fixed. Resulting eclipse time 
variation diagram is shown in Figure~\ref{Fig_etv}. In the 
figure, irregular undulating variation is noticeable among 
the scatter, with an approximate amplitude of 0\fd003. Since 
that pattern does not repeat itself strictly during four years, 
it is not likely to attribute it to a possible third body. 
More likely explanation of this pattern might be the out-of-eclipse 
variability of the system, which can be notice in b and c 
panels of Figure~\ref{Fig_all_LC}.

\subsection{Radial velocities and spectroscopic orbit}\label{Sec_spec_orb}

\begin{figure}
	\includegraphics[scale=0.73]{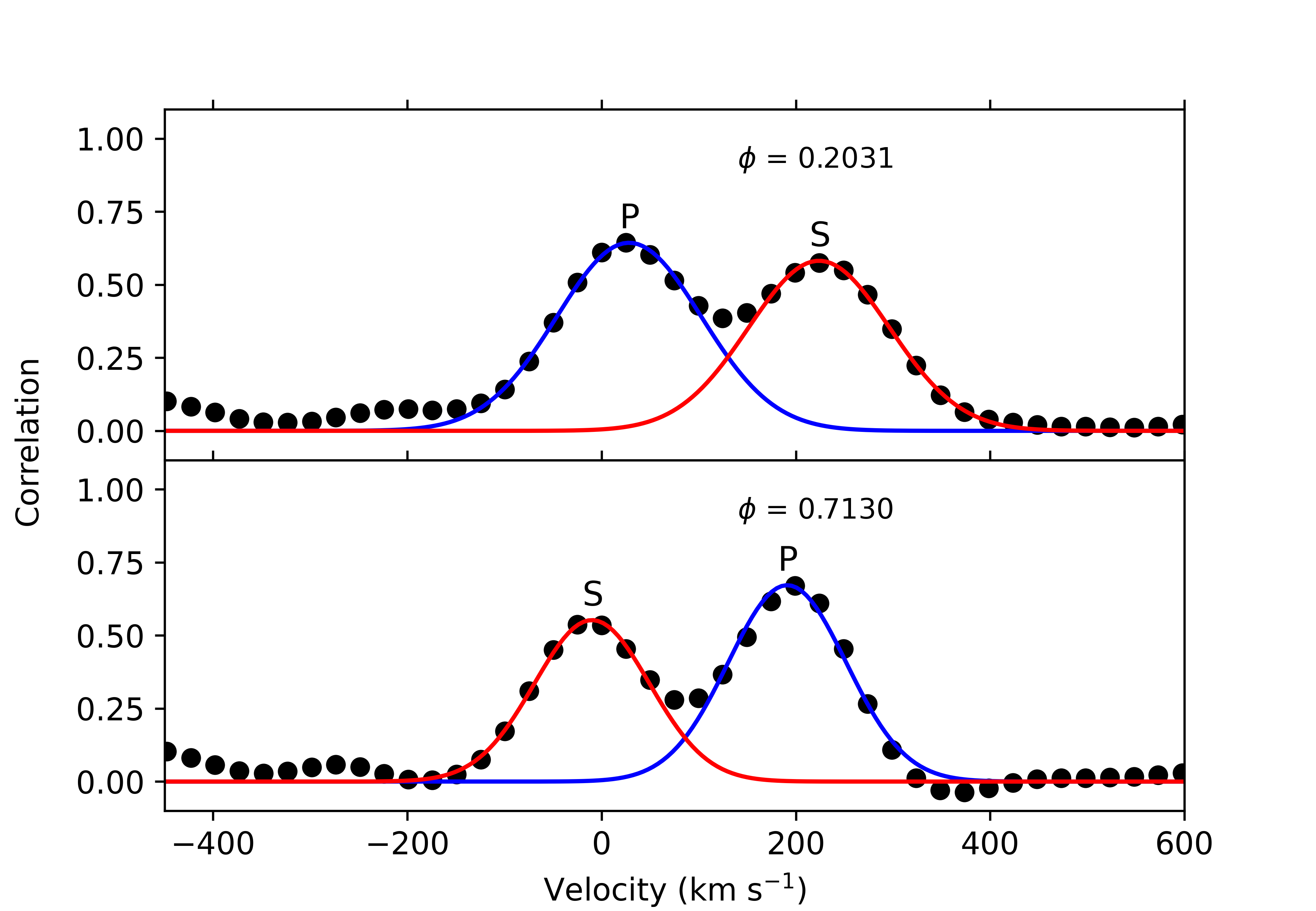}
	\caption{Cross-correlation functions of two observed 
	spectra recorded around orbital quadratures. P and S 
	denote the primary and the secondary component, 
	$\phi$ shows orbital phase.}
    \label{Fig_ccf}
\end{figure}

Next step of our analysis is to determine radial velocities
of each component from each observed spectrum and model the 
spectroscopic orbit of the system. In order to determine
radial velocities of the components, we use optical spectrum 
of HD\,184499 \citep[$T_{eff}$ = 5743, log $g$ = 4.07;][]{hd184499_Pruniel_et_al_2011A&A...531A.165P} 
as a template spectrum, which was recorded with the same 
instrumental set-up at night of 20th August, 2014. Then, 
each observed spectrum of KIC\,4832197 was cross-correlated 
with the spectrum of HD\,184499 by following the method 
proposed by \citet{fxcor_Tonry_Davis_1979}. Practical 
application of this method was done by $fxcor$ task 
\citep{fxcor_Fitzpatrick_1993ASPC} under IRAF environment. 
All clear absorption lines, except strongly blended or very 
broad spectral lines, were considered in 5th and 6th \'echelle 
orders, which cover wavelength range between 4900 \AA~ and 
5700 \AA~. In this wavelength range, it was possible to 
detect strong signals of both components. We show 
cross-correlation functions of two spectra of 
KIC\,4832197 recorded at two separate orbital quadratures, 
in Figure~\ref{Fig_ccf}. Determined heliocentric radial 
velocities are tabulated in Table~\ref{Table_rv}.

\begin{table}
\setlength{\tabcolsep}{3pt}
\small
\caption{Summary of spectroscopic observations along with 
measured radial velocities and their corresponding standard 
errors ($\sigma$) in \kms. SNR denotes signa-to-noise ratio 
around 5500\AA wavelength.}\label{Table_rv}
\begin{center}
\begin{tabular}{ccccrrrr}
\hline\noalign{\smallskip}
      HJD    & Orbital  & Exposure & SNR &\multicolumn{2}{c}{Primary} &  \multicolumn{2}{c}{Secondary} \\
(24 00000+)  &  Phase   & time (s) &     & V$_{r}$ & $\sigma$ & V$_{r}$ & $\sigma$  \\
\hline\noalign{\smallskip}
56842.4581	&	0.7936	&	3200	&	120	&	61	&	10	&	-150	&	11	\\
56842.4963	&	0.8138	&	3200	&	95	&	58	&	9	&	-146	&	10	\\
56843.3937	&	0.2872	&	3200	&	90	&	-141	&	10	&	70	&	15	\\
56887.3009	&	0.4516	&	3200	&	140	&	-79	&	9	&	-8	&	11	\\
56887.5063	&	0.5599	&	3200	&	95	&	7	&	11	&	-74	&	13	\\
56888.3307	&	0.9949	&	3200	&	120	&	-43	&	6	&	---	&	---	\\
57592.3910	&	0.4395	&	3600	&	90	&	-81	&	9	&	-1	&	9	\\
57600.4912	&	0.7130	&	3600	&	110	&	59	&	11	&	-154	&	13	\\
57617.3447	&	0.6045	&	3600	&	90	&	17	&	10	&	-106	&	17	\\
57853.5170	&	0.2031	&	2700	&	100	&	-135	&	13	&	72	&	16	\\
\noalign{\smallskip}\hline
\end{tabular}
\end{center}
\end{table}

\begin{figure}
	\includegraphics[scale=0.73]{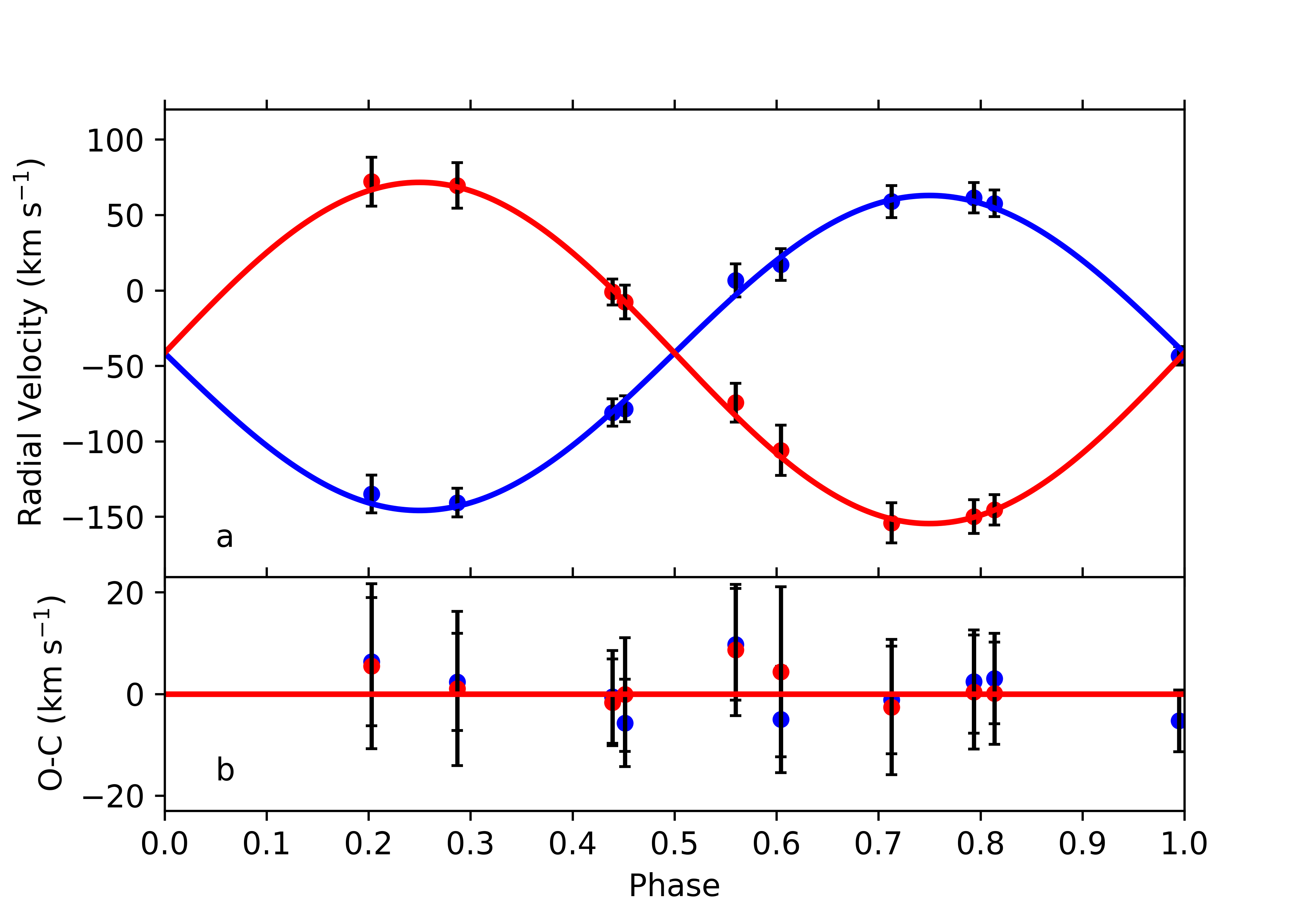}
	\caption{Phase-folded radial velocities of the primary 
	(blue) and the secondary (red) components are shown as 
	filled circles. The best-fitting spectroscopic orbit 
	model is over-plotted for each component as continuous 
	line.}
    \label{Fig_rv_model}
\end{figure}

Preliminary inspection of the phase-folded light curve 
shows that the mid-primary and the mid-secondary eclipses 
precisely occur at 0.0 and 0.5 orbital phases, respectively. 
This finding strongly suggests circular orbit. Therefore, 
spectroscopic orbit of the system is determined under 
zero eccentricity (i.e. $e=0$) assumption. In this case, 
longitude of the periastron ($\omega$) is undefined, thus 
$T_{0}$ value found in light time variation analysis step 
is adopted instead of periastron passage time. $T_{0}$
and $P$ values are kept fixed during the modelling, while
radial velocity semi-amplitudes of the components 
($K_{1}$ and $K_{2}$), center-of-mass velocity of the 
system ($V_{\gamma}$) are adjusted. Application of linear 
least squares fitting method to the observed radial 
velocities results in the best-fitting spectroscopic orbit 
parameters tabulated in Table~\ref{Table_spec_orbit}. 
Agreement between observed radial velocities and 
best-fitting model is shown in Figure~\ref{Fig_rv_model}.

\begin{table}
\caption{Best-fitting spectroscopic orbit parameters of 
KIC\,4832197. $M{_1}$ and $M{_2}$ denote the masses of 
the primary and the secondary component, respectively, 
while $M$ shows the total mass of the system and 
$a\sin i$ is the projected semi-major axis, depending on 
the orbital inclination $i$.}\label{Table_spec_orbit}
\begin{center}
\begin{tabular}{cc}
\hline\noalign{\smallskip}
Parameter & Value \\
\hline\noalign{\smallskip}
$P_{\rm orb}$ (day)     &    1.8954655 (fixed)  \\
$T_{\rm 0}$ (HJD24 00000+)   &  54954.965798 (fixed)  \\
$V_{\gamma}$ (\kms)          &    $-$41$\pm$2     \\
$K_{1}$ (\kms)           &     104$\pm$4     \\
$K_{2}$ (\kms)           &    113$\pm$5     \\
$e$                      &      0 (fixed)       \\
$a\sin i$ (\Rsun)        &     8.15$\pm$0.26    \\
$M\sin^{3} i$ (\Msun)    &    2.03$\pm$0.14  \\
Mass ratio ($q = M{_2}/M{_1}$)         &     0.92$\pm$0.06    \\
rms1 (\kms )           &         5        \\
rms2 (\kms )           &         4         \\
\noalign{\smallskip}\hline
\end{tabular}
\end{center}
\end{table}

\subsection{Spectral type}\label{Sec_spec_type}

Medium resolution TFOSC spectra allow to determine spectral 
types and global atmospheric parameters of each component 
of KIC\,4832197. A spectrum recorded around any 
of the orbital quadrature is fairly suitable for atmospheric 
parameter determination of individual components. Because, 
spectral lines of both components are often separated from 
each other along wavelength axis sufficiently and can be 
distinguished at those orbital phases. This is exactly 
observed in TFOSC spectrum recorded at 30th July 2016 night, 
corresponding 0.7130 orbital phase (see lower panel of the 
Figure~\ref{Fig_ccf}). Spectral types and atmospheric 
parameters are determined from this spectrum.

Determination of spectral types and atmospheric parameters 
is an iterated process as in the case of determination of 
mid-eclipse times described in Section~\ref{Sec_analysis_light_time}. 
During analysis, we fix micro-turbulence velocity of each 
component to 2 \kms. Actually, analysing high resolution optical 
spectra, it is possible to determine micro-turbulence velocity 
and [Fe/H] abundance simultaneously via Blackwell diagram 
\citep{Blackwell_diagram_1979MNRAS.186..673B}. However, due to 
the insufficient resolution of TFOSC spectra, we are unable to 
do so. Instead, we implicitly assume 2 km/s micro-turbulence 
velocity and fix this value during analysis. Although there 
isn't any strict relation in the literature to estimate 
micro-turbulence velocities reliably, limited observational 
studies indicate that 2 \kms of micro-turbulence velocity is 
proper for solar type stars 
\citep{microturbulence_ref_2009A&A...503..973L}.

It is possible to apply a constraint on logarithm of 
gravity (log $g$) values of the components. To do this, 
spectroscopic orbit model parameters are combined with 
preliminary light curve solution parameters. This step 
allows computation of masses and radii of the components. 
Then, log $g$ values are computed via computed masses 
and radii of the components. Computed log $g$ values are 
fixed in spectral type analysis and effective temperatures 
of the components ($T_{eff1}$ and $T_{eff2}$) are 
estimated together with overall metallicity ($[Fe/H]$). 
Then, estimated temperature of the primary component and 
overall metallicity are adopted as fixed parameters and 
combined light curve and radial velocity modelling is 
repeated. Two or three iteration is fairly enough to 
reach a self-consistent solution. 

Effective temperatures and overall metallicity are 
estimated by spectrum synthesizing method. ATLAS9 
\citep{ATLAS9_castelli_2004} model atmospheres, which 
adopts plane-parallel atmosphere assumption, are used for 
synthetic spectrum computation. A grid of synthetic 
spectra are computed for temperature range of 6000 K and 
7000 K with a step of 100 K. This computation is repeated 
for metallicities between solar ($[Fe/H]=0$) and sub-solar 
($[Fe/H]=-1.0$), with a step of 0.25. Practical computation 
of synthetic spectra are done by a \textsc{Python} 
framework, $iSpec$ software \citep{iSpec_Cuaresma_2014A&A}. 
Among various radiative transfer codes provided in $iSpec$, 
\textsc{Spectrum} code \citep{spectrum_gray_1994} is 
adopted. $iSpec$ also includes a comprehensive line list 
compiled from third version of the Vienna atomic line 
database \citep[VALD3,][]{VALD3_Ryabchikova_2015}. Each 
computed synthetic spectrum is convolved with a Gaussian 
line spread function to reduce spectral resolution to the
resolution of TFOSC spectra. After that point, a trial
synthetic spectrum is chosen for each component among 
computed spectra and composite spectrum of the system is 
computed by those spectra. In order to compute composite 
spectrum of the system, each individual spectrum is 
shifted in wavelength axis with respect to the radial 
velocity of the corresponding component and scaled with 
respect to the square of ratio of the radii of the 
components.

\begin{figure}
	\includegraphics[scale=0.73]{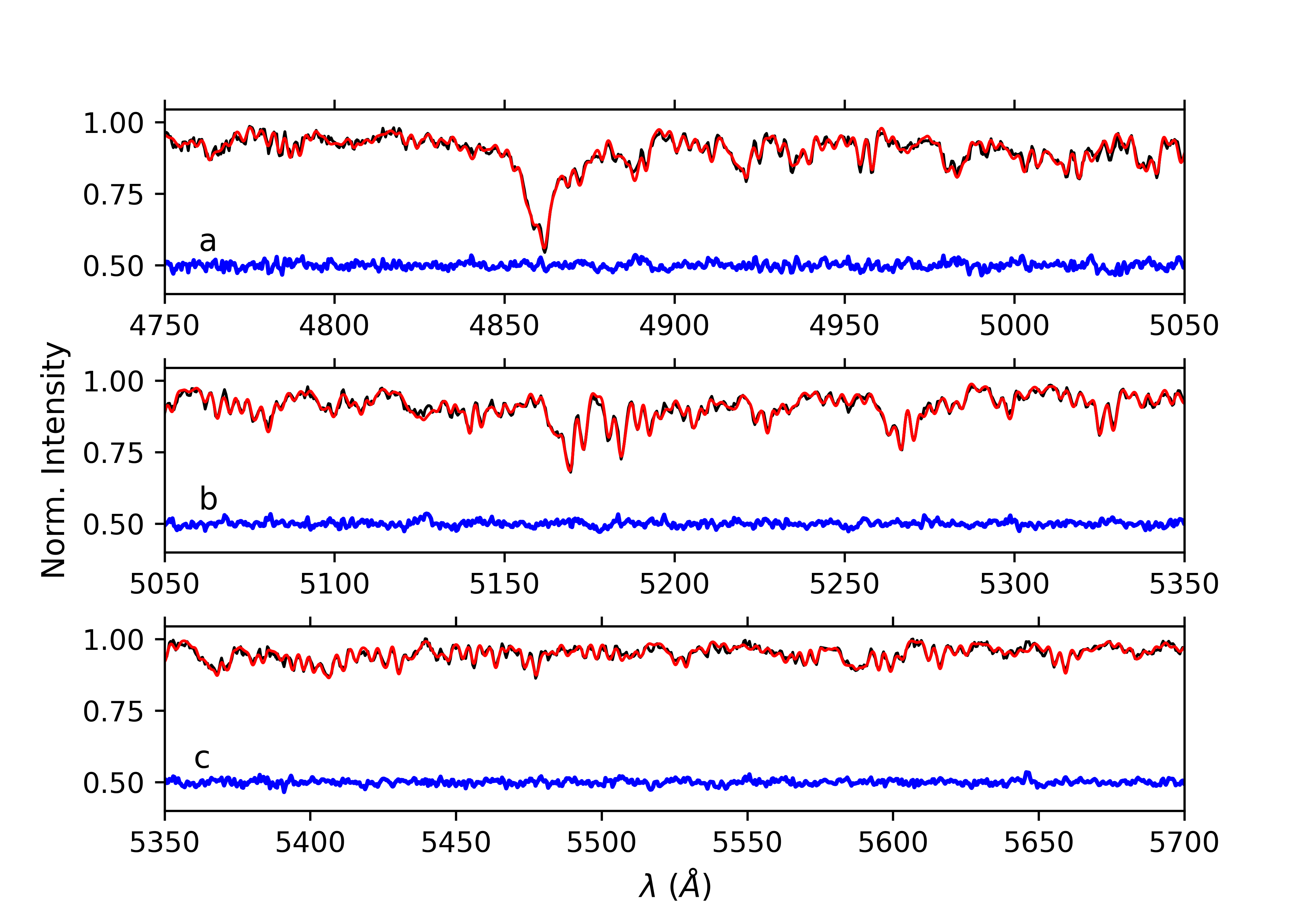}
	\caption{Observed (blue) and computed composite 
    spectrum (red) for KIC\,4832197 in three different 
    portions of the optical wavelengths. Residuals from 
    the best-fitting model are plotted as shifted upwards 
    by 0.5 (blue) for a better viewing purpose.}
    \label{Fig_spectrum}
\end{figure}

Self-consistent effective temperatures and overall 
metallicity are determined in the third iteration, which 
gives $T_{eff1}$ = 6300 K, $T_{eff2}$ = 6100 K and 
[Fe/H] = $-$0.25. Estimated uncertainties are around
200 K for temperatures and 0.25 for metallicity. Adopted
ratio of radii is $R_{1}/R_{2}=1.27$ and log $g$ values 
are log $g_{1}=4.30$ and log $g_{2}=4.47$ (see next 
section for details). These results indicate F7V and F9V
spectral types for the primary and the secondary 
components \citep{Gray_2005}, respectively. Observed 
TFOSC spectrum and best-fitting composite spectrum are 
shown in Figure~\ref{Fig_spectrum}.

\subsection{Light curve analysis and evolutionary status}\label{Sec_lc_analysis_evol_stat}

Long cadence photometric data of KIC\,4832197 include
65190 individual data points in total, which means large 
amount of CPU time is required to find a best-fitting 
light curve model. Therefore, an average phase-folded 
light curve is prepared by computing average in every 
0.01 phase at out-of eclipse phases. Around mid-eclipse
phases, phase step is adopted as 0.0005 in order to
detect ingress and egress phases of eclipses precisely.
Average phase-folded light curve includes 556 data 
points in total, which enormously reduces required 
CPU time for light curve modelling. Weighting of each 
normal data point in phase-folded average light curve
is done by considering total number of data points, 
which produces the normal data point.

Light curve modelling is done by 2015 version of the
well-known Wilson--Devinney (WD) eclipsing binary light 
curve modelling code \citep{WD_MAIN_1971ApJ, WD2015_2014ApJ}.
Practical application with WD code is done by user 
friendly \textsc{Python} GUI $PyWD2015$ 
\citep{PyWD2015_2020CoSka..50..535G}. $PyWD2015$ allows 
users to use almost all features of the 2015 version of 
the WD code, expect subsets. Furthermore, in addition to
capabilities of the WD code, $PyWD2015$ includes many 
useful features and small tools to speed up modelling 
process and tracing successive iterations visually.

\begin{figure}
	\includegraphics[scale=0.73]{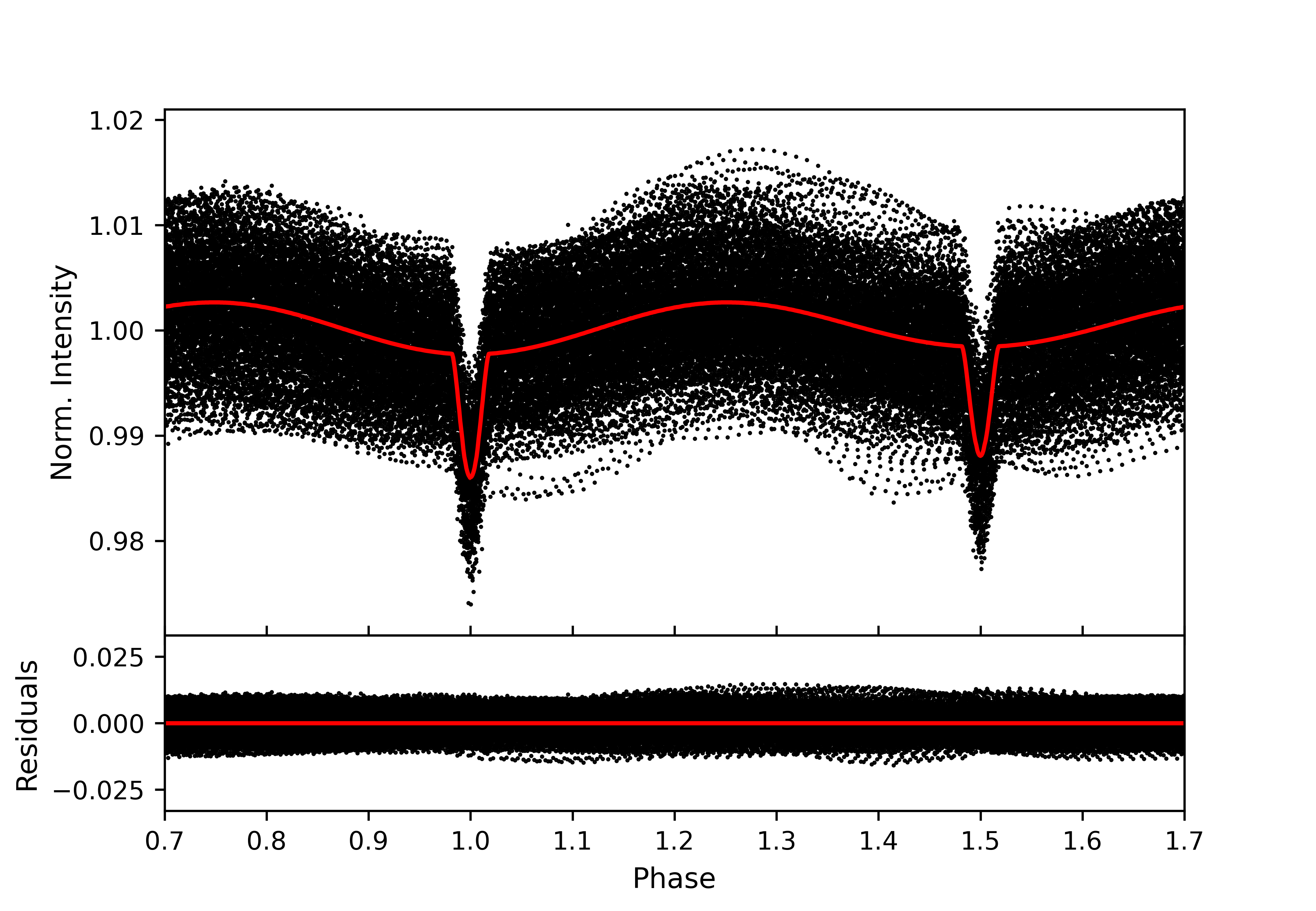}
	\caption{Phase-folded long cadence Kepler observations
	and the best-fitting light curve model of KIC\,4832197
	are shown along with residuals from the model.}
    \label{Fig_lc_obs_model}
\end{figure}

\begin{figure}
	\includegraphics[scale=0.73]{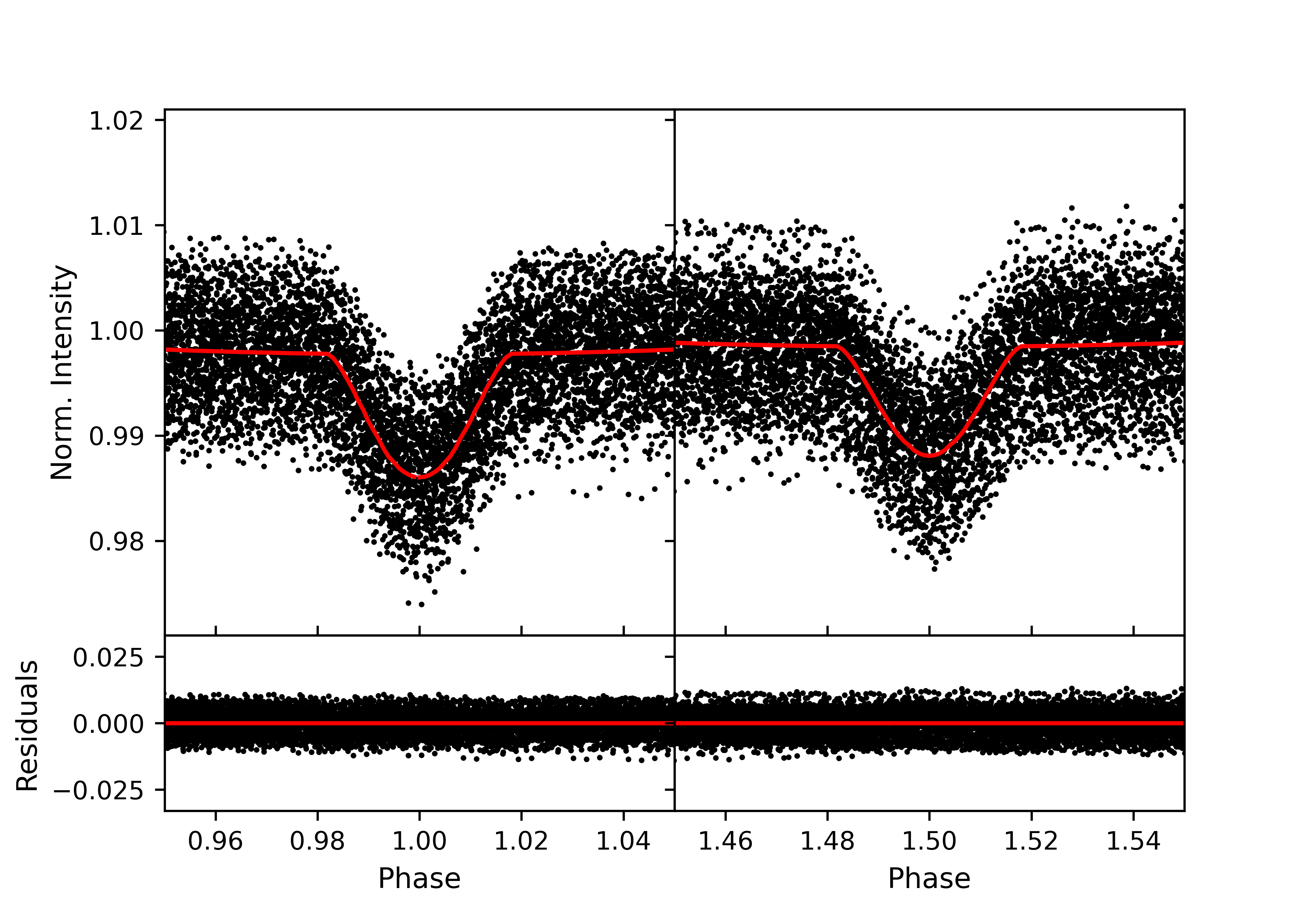}
	\caption{Close-up view of observations and the 
	best-fitting light curve model around eclipse phases.}
    \label{Fig_lc_obs_model_eclipses}
\end{figure}

Two most critical parameters for an accurate light curve
modelling are $T_{eff1}$ and $q$. These have already been
determined in previous sections, are kept fixed during
modelling. Effective temperatures of the components
clearly indicate that both components possess convective 
outer envelope, thus gravity darkening ($g$) and albedo
($A$) values of each component are set to 0.32 
\citep{Lucy_1967ZA.....65...89L} and 0.5 
\citep{Rucinski_albedo_1969AcA....19..245R}, 
respectively. Low resolution of TFOSC spectra does not
allow to determine reliable rotational velocities of the
components, so considering non-eccentric orbit of the 
system, synchronous rotation is implicitly assumed for 
both components by fixing rotation parameter ($F$) of 
each component to the unity. Logarithmic limb darkening 
law \citep{limb_dark_log_Klinglesmith_1970AJ} is adopted 
for both components and limb darkening coefficients are 
adopted from \citet{van_Hamme_LD_1993AJ} for Kepler 
passband. Adjustable parameters are inclination of the 
orbit ($i$), effective temperature of the secondary 
component ($T_{eff2}$), dimensionless potentials of the 
components ($\Omega_{1}$, $\Omega_{2}$) and luminosity of 
the primary component ($L_{1}$). Coarse and fine grid numbers 
for stellar surface are set to 60. Modelling attempts with 
a possible third light contribution do not yield a 
reasonable model, which mostly indicate extremely small 
or negative third light contribution, therefore modelling 
process is carried out with no third light contribution. 
Convergence is not very quick due to the shallow eclipse 
depths, however, successive iterations 
converge slow but firmly to a global minimum. Model 
parameters for the best-fitting light curve are tabulated 
in Table~\ref{Table_lc_solution}. 
In Figure~\ref{Fig_lc_obs_model}, phase-folded light 
curve and and best-fitting light curve model are shown. 
Separate close-up view of each eclipse is shown in 
Figure~\ref{Fig_lc_obs_model_eclipses}.

\begin{table}[!htb]
\caption{Best-fitting light curve model parameters 
for KIC\,4832197. $\langle r_{1}\rangle$ and 
$\langle r_{2}\rangle$ show mean fractional radii of 
the primary and the secondary components, respectively. 
Uncertainty of adjusted parameters are internal to 
the Wilson-Devinney code and given in parentheses 
for the last digits, except $T_{2}$. Uncertainty of 
$T_{2}$ is assumed to be the same as $T_{1}$.}\label{Table_lc_solution}
\begin{center}
\begin{tabular}{cc}
\hline\noalign{\smallskip}
Parameter & Value \\
\hline\noalign{\smallskip}
$q$ &  0.92 (fixed) \\
$T_{1}(K)$ &  6500 (fixed) \\
$g_{1}$ = $g_{2}$  & 0.32\\
$A_{1}$ = $A_{2}$  & 0.5\\
$F_{1}$ = $F_{2}$  & 1.0 \\
$i~(^{\circ})$ &  75.67(5)\\
$T_{2}(K)$ &  6060(200)\\
$\Omega_{1}$ & 7.605(22)\\
$\Omega_{2}$ & 8.520(44) \\
$L_{1}$/($L_{1}$+$L_{2})$ & 0.639(3) \\
$x{_1}, x{_2}$  & 0.685, 0.700 \\
$y{_1}, y{_2}$  & 0.270, 0.259 \\
$\langle r_{1}\rangle, \langle r_{2}\rangle$ & 0.1498(5), 0.1219(8) \\
Model rms           &     2.8 $\times$ 10$^{-4}$   \\
\noalign{\smallskip}\hline
\end{tabular}
\end{center}
\end{table}

Combination of the best-fitting spectroscopic orbit and 
light curve model parameters allows to compute absolute
dimension of the system as well as its distance. Both
$UBV$ colours \citep{ubv_kepler_2012PASP..124..316E} and 
optical spectrum of the system indicate F7 spectral type, 
which means reddening and interstellar extinction should 
be negligible, if any. Trial plotting of the system on
$UBV$ colour-colour diagram gives $E(B-V)=0\fm009$, which
is pretty below the observational error of $B-V$, hence
might be ignored for distance computation. Considering 
$P$, $K_{1}$, $K_{2}$, $i$, $\langle r_{1}\rangle$ and 
$\langle r_{2}\rangle$ along with the published $UBV$ 
magnitudes, absolute dimension of the system is 
computed by \textsc{JKTABSDIM} 
code\footnote{https://www.astro.keele.ac.uk/jkt/codes/jktabsdim.html} 
\citep{jktabsdim_southworth_et_al_2005A&A...429..645S}
and tabulated in Table~\ref{Table_abs_dim}. Calibrations
given in \citet{kervella_et_al_2004A&A...426..297K} 
is adopted for bolometric correction and distance 
computation.

\begin{table}
\caption{Absolute dimension of KIC\,4832197. Statistical 
error of each parameter is given in parentheses for the 
last digits.}\label{Table_abs_dim}
\begin{center}
\begin{tabular}{ccc}
\hline\noalign{\smallskip}
Parameter & Primary & Secondary \\
\hline\noalign{\smallskip}
Spectral Type      &  F7V     &  F9V  \\
Mass (\Msun)       &  1.16(12) & 1.07(10) \\
Radius (\Rsun)     &  1.26(4) & 1.03(3) \\
Log $L/L_{\odot}$   & 0.35(6) & 0.11(6) \\
log $g$ (cgs)      &  4.30(2) & 4.44(2) \\
$M_{bol}$ (mag)      & 3.87(15) & 4.48(16) \\
\multicolumn{1}{c}{[Fe/H]} & \multicolumn{2}{c}{$-$0.25(0.25)} \\
\multicolumn{1}{c}{separation (\Rsun)} & \multicolumn{2}{c}{8.41(26)} \\
\multicolumn{1}{c}{distance (pc)} & \multicolumn{2}{c}{459(40)} \\
\noalign{\smallskip}\hline
\end{tabular}
\end{center}
\end{table}

\begin{figure}
    \includegraphics[scale=0.73]{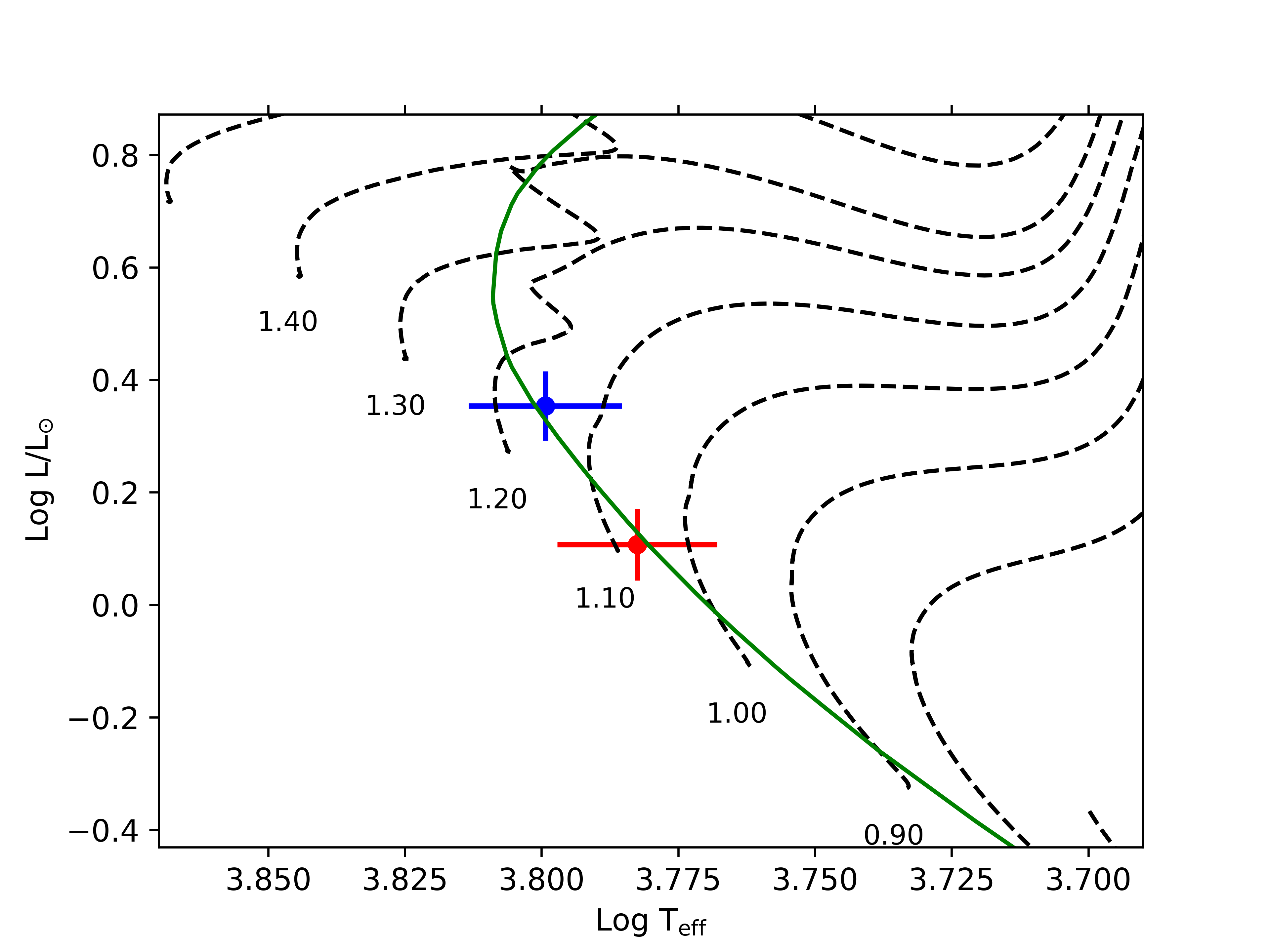}
    \caption{Positions of the primary and the secondary 
        components on $Log~T_{eff}-Log~L/L_{\odot}$ plane. 
        Dashed lines show evolutionary tracks for different 
        masses. Each track is labelled with its corresponding 
        mass. Continuous curve shows the isochrone for 
        log(age)=9.45. The tracks are for slightly sub-solar 
        metallicity with $Y = 0.273$ and $Z = 0.014$.}
    \label{Fig_hr_diagram}
\end{figure}

Comparing the locations of the components with stellar 
evolutionary tracks on $Log~T_{eff}-Log~L/L_{\odot}$ 
plane, it is possible to estimate the age of the system. 
In Figure~\ref{Fig_hr_diagram}, components of KIC\,4832197 
are plotted along with stellar evolutionary tracks 
computed by PAdova and TRieste Stellar Evolution 
Code \citep[PARSEC,][]{bressan_et_al_2012MNRAS.427..127B}. 
The best-matching isochrone to the positions of the 
components has log(age)=9.45, which indicates 2.8 Gyr 
age for the system. Estimated uncertainty for this 
age is approximately 0.8 Gyr. Primary source of this 
uncertainty is the uncertainty of estimated effective 
temperatures and its propagation on computed 
luminosities. Both components are still on the main 
sequence, however, the primary component is almost 
half-way through its main sequence life.

\subsection{Out-of-eclipse variability}\label{Sec_lc_out_of_eclipses}

\begin{figure}
	\includegraphics[scale=0.73]{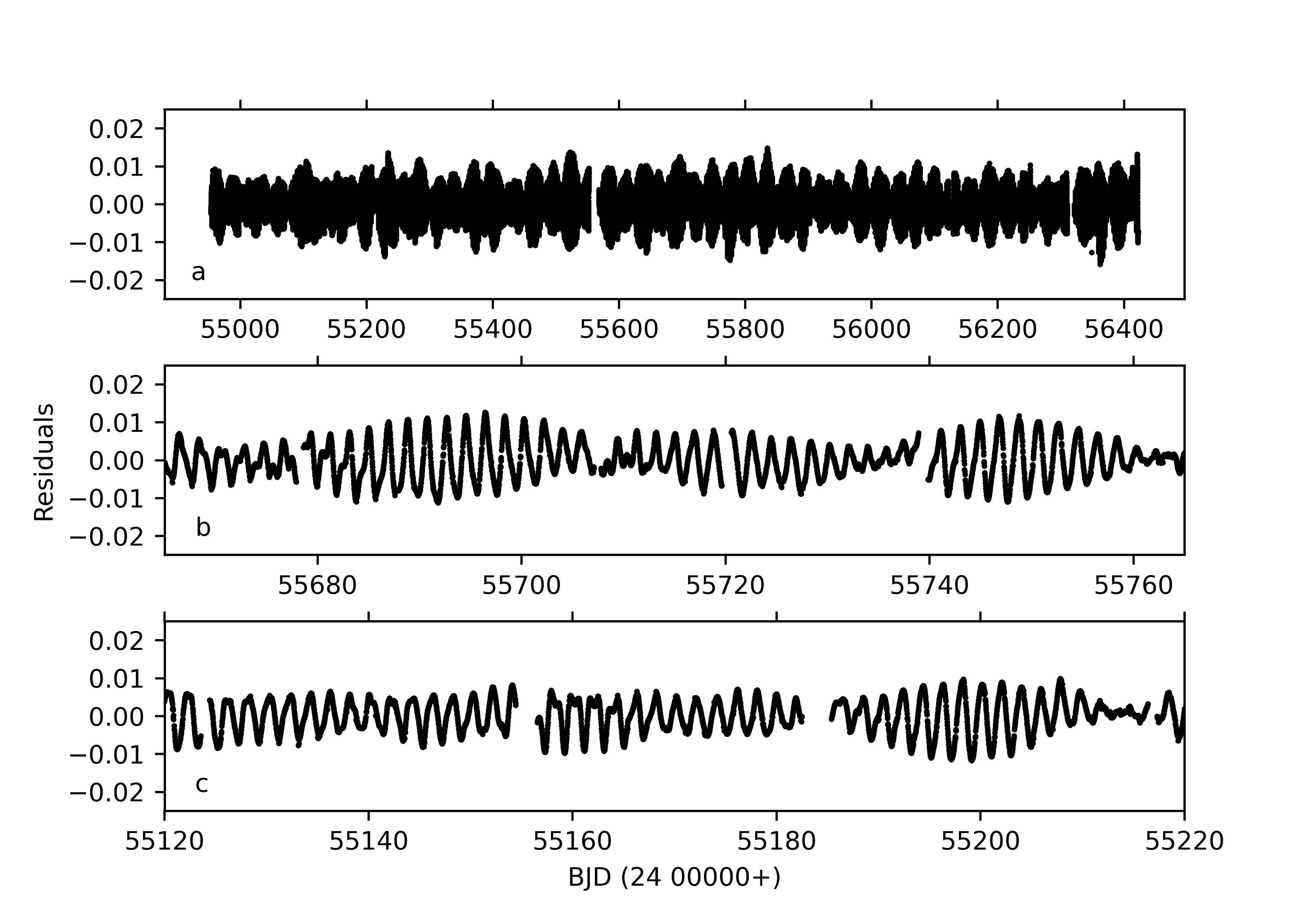}
	\caption{Residuals from the best-fitting light curve 
	model are plotted against time ($BJD$). Panel $a$ shows 
	the whole long cadence residuals, while panels $b$ and
	$c$ show different portions of the whole set.}
    \label{Fig_lc_residuals}
\end{figure}

A scatter of $\pm0.01$ is clearly noticeable in phase-folded 
light curve (see Figure~\ref{Fig_lc_obs_model_eclipses}). 
This scatter is natural result of dominant 
brightness variability at out-of-eclipse phases 
(see Figure~\ref{Fig_all_LC}, lower panels). For 
further investigation on this variability, best-fitting 
light curve model is subtracted from observations 
and residuals from the model are obtained. 
In principle, these residuals do not include variability 
due to eclipses and proximity effects of the components 
and give hints on intrinsic variability of one or both 
components. Residuals from the best-fitting light curve 
model is obtained with \textsc{PyWD2015} by inputting all long 
cadence data into WD code and make a single differential 
corrections iteration. Obtained residuals are plotted in 
Figure~\ref{Fig_lc_residuals}. Quick changes in residual 
light curve shape and amplitude are remarkable. 

\begin{figure}
	\includegraphics[scale=0.73]{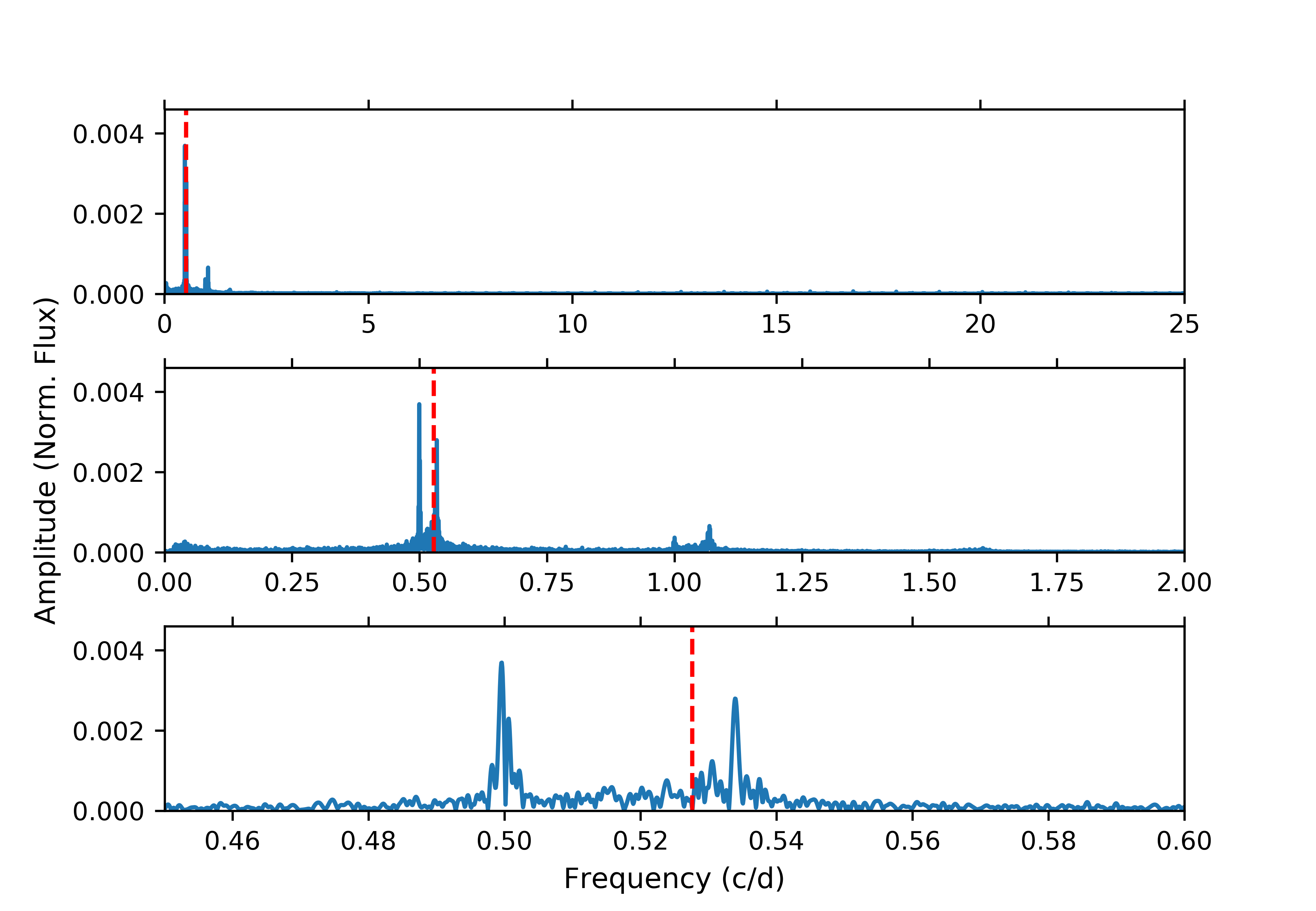}
	\caption{Amplitude spectrum of light curve residuals. 
	The uppermost panel show the whole amplitude spectrum 
	between 0 and 24.498 c/d. Middle panel focuses on 0 and 2 
	c/d where dominant peaks appear. Lowermost panel shows 
	detailed view of the most dominant two peaks and the 
	orbital frequency. In each panel, the orbital frequency
	is marked with a vertical dashed (red) line.}
    \label{Fig_periodogram}
\end{figure}

In order to see the reflection of this variability in 
frequency-amplitude plane, Lomb-Scargle periodogram 
\citep{Lomb_1976Ap&SS, Scargle_1982ApJ} is applied to whole 
long cadence residuals via python script \textsc{pysca} 
\citep{pysca2014..301..421H}. \textsc{pysca} allows quick 
and practical computation of amplitude spectrum and extract 
significant frequencies above a defined noise level. Adopting
24.498 cycle/day (c/d) as nyquist frequency, single 
\textsc{pysca} run results in the amplitude spectrum given 
in Figure~\ref{Fig_periodogram}. Adopted nyquist frequency 
corresponds to the twice ($\approx$ 59 min) of integration 
time of a single long cadence exposure. In the resulting
amplitude spectrum, two dominant peaks are clearly seen around 
the orbital frequency but no signal appear at precise location
of the orbital frequency. Frequency values of the two dominant 
peaks are tabulated in Table~\ref{Table_frequencies}, 
together with the orbital frequency.

\begin{table}[!htb]
\caption{First two dominant frequencies ($f_{1}$, $f_{2}$) 
resulting from frequency analysis of residual light curve 
of KIC\,4832197 along with orbital frequency ($f_{orb}$) of 
the system. Uncertainties are given in parentheses for the 
last digits.}\label{Table_frequencies}
\begin{center}
\begin{tabular}{cccc}
\hline\noalign{\smallskip}
Frequency	&	Frequency 	&	     Amplitude              &	 Phase 		\\
number		&	(c/d)		&	(Flux $\times10^{3}$)		&	(radian)	\\
\hline\noalign{\smallskip}									
$f_{1}$		&	0.4995769(76)	&	3.70(2)	&	0.4947(8)	\\
$f_{2}$		&	0.5339561(88)	&	2.79(2)	&	0.807(1)	\\
$f_{orb}$	&	0.5275750(1)	&	---		&	  ---		\\
\noalign{\smallskip}\hline

\end{tabular}
\end{center}
\end{table}

\begin{figure}
    \includegraphics[scale=0.63]{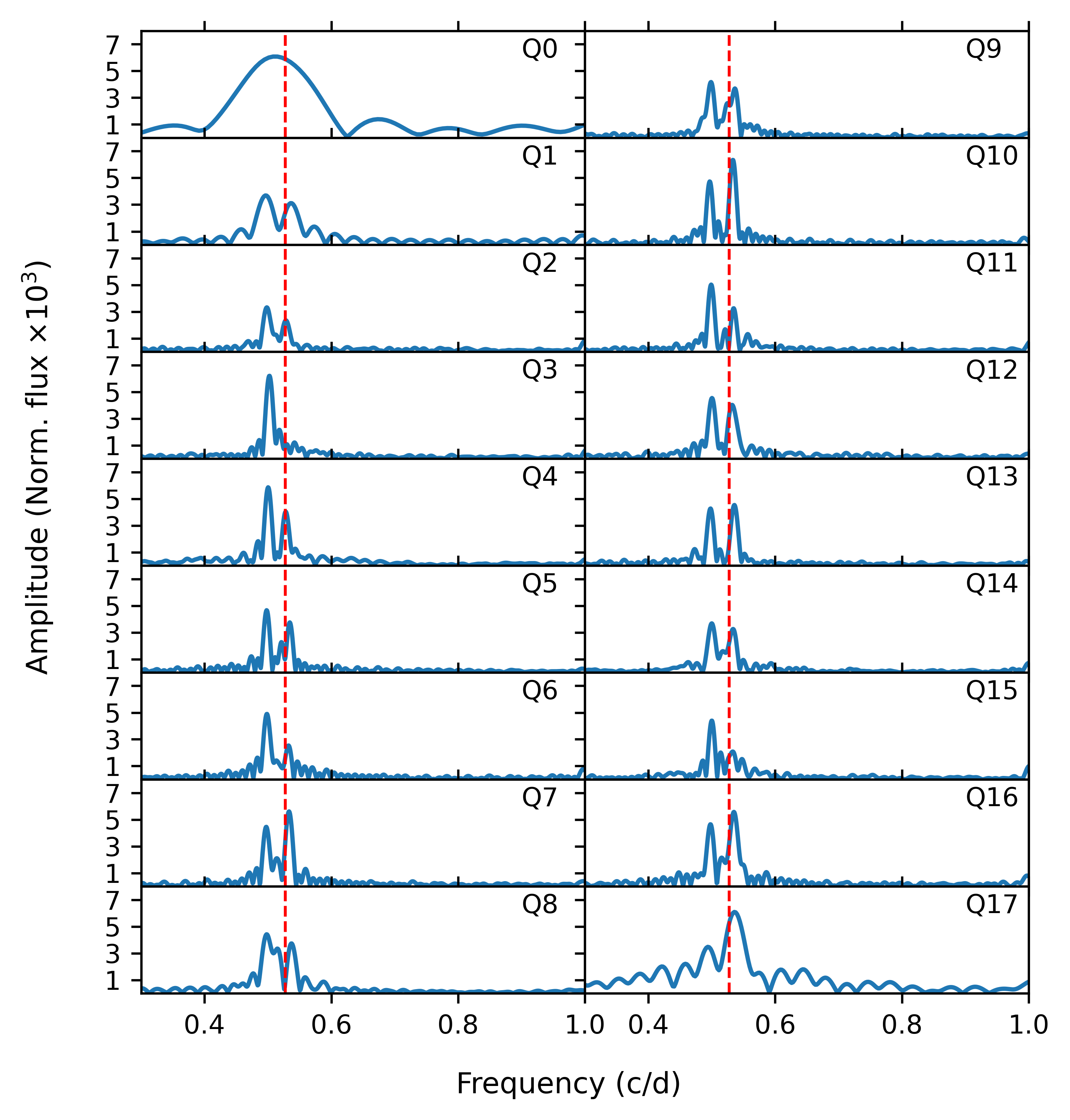}
    \caption{Close view of amplitude spectrum of light curve 
        residuals for each quarter of Kepler data. Panels 
        are focused on frequency region around $f_{orb}$, 
        which is shown by vertical (red) dashed line. Each 
        panel is labelled with its corresponding quarter in 
        upper left corner.}
    \label{Fig_periodogram_quarters}
\end{figure}

\begin{figure}
    \includegraphics[scale=0.73]{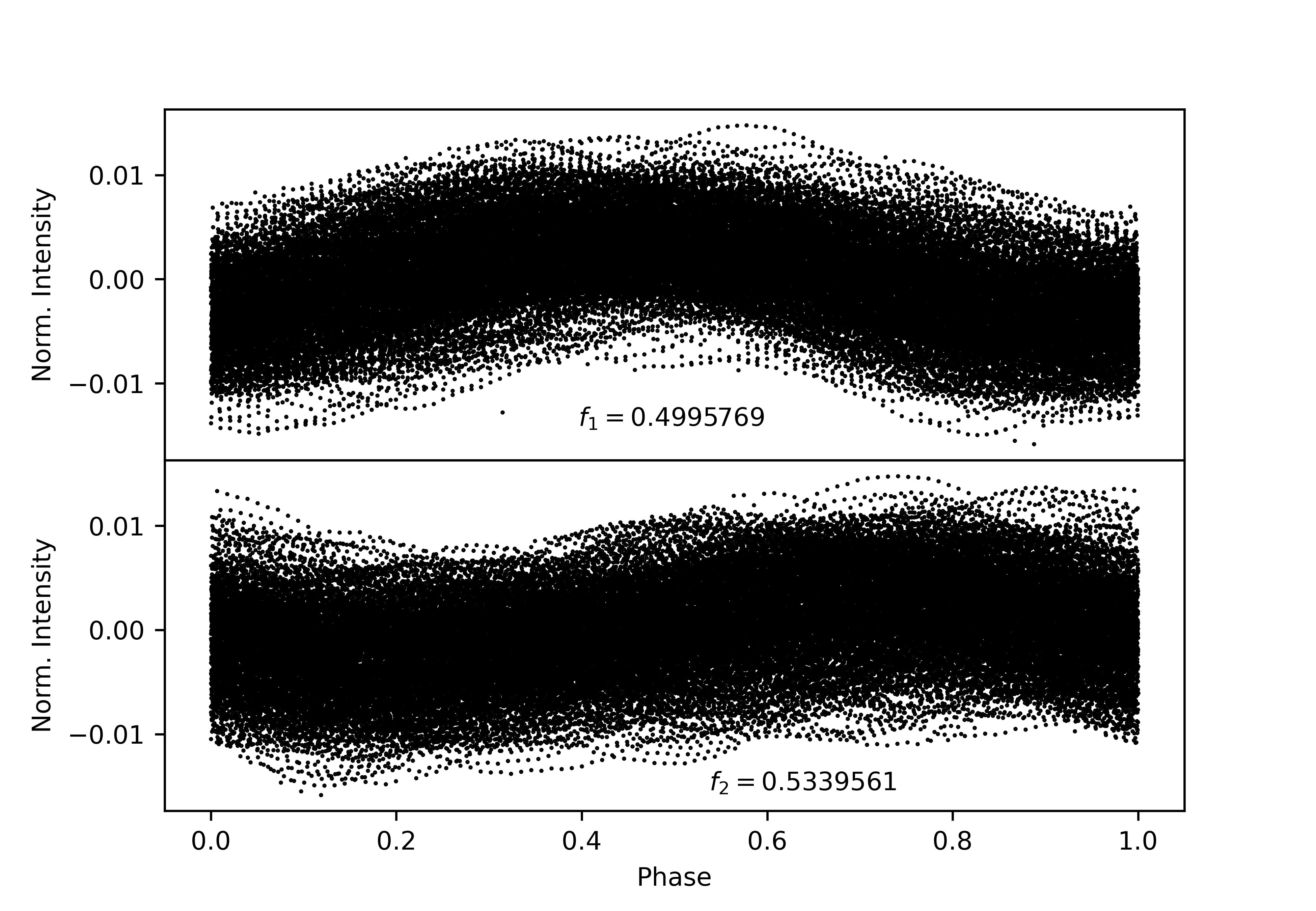}
    \caption{Phase folded light curve residuals with respect to
        $f_{1}$ and $f_{2}$ frequencies. Each frequency is shown in
        corresponding panel.}
    \label{Fig_f1_f2_phased_lcs}
\end{figure}

It is also possible to trace the behaviors of the detected 
frequency peaks for each Kepler quarter separately. Close view 
of the $f_{1}$ and $f_{2}$ peaks are shown in 
Figure~\ref{Fig_periodogram_quarters} for each Kepler quarter,
separately. It is apparent that $f_{1}$ strongly appears in most 
of quarters while $f_{2}$ weakens or disappears in some quarters.

The corresponding frequencies of the first two dominant peaks 
in the amplitude spectrum give the phase folded residual 
light curves shown in Figure~\ref{Fig_f1_f2_phased_lcs}. Phase
folding is done with respect to the reference ephemeris given in
Equation~\ref{Eq2} and corresponding period values of the frequencies
listed in Table~\ref{Table_frequencies}.
Wave-like light curve pattern can be noticed easily in both 
panels of the figure. This picture appears to be the reflection 
of double-humped structure of the light curve 
(see Figure~\ref{Fig_all_LC}, panel $b$ and $c$), which is 
frequently observed through four years of long cadence data.
Estimated spectral types of the components in 
Section~\ref{Sec_spec_type} suggests that these $f_{1}$ and 
$f_{2}$ frequencies are likely the result of cool spot activity 
on one or both components.

\section{Summary and Discussion}\label{Sec_summary_discussion}

Analyses of medium resolution ground-based optical spectroscopy 
and very high precision Kepler photometry show that KIC\,4832197 
is composed of F7V and F9V stars possessing slightly sub-solar 
metallicity ([Fe/H]=$-$0.25$\pm$0.25). The components moves
around the center-of-mass of the system on a circular orbit with 
a period of $1\fd8954650$. Kepler long cadence light curve of the system
is dominated by a wave-like brightness variability whose amplitude
is comparable to the eclipse depths. In some epochs, amplitude
of this variability exceeds the amplitude of eclipses, which are
very shallow and hardly exceeds 0.01 in normalized flux unit.

Shallow eclipse depths could be result of a possible third light 
contribution, however, the best-fitting light curve model indicates
no third light contribution. ETV diagram does not show any sign for
a possible third body, but only a scatter of $\pm$ 5 minute with some
vaguely undulating patterns among this scatter. Remaining possible
explanation is that the undulating pattern should be the reflection
of the variability observed at out-of-eclipse phases. Considering 
spectral types of the components, such irregular patterns in etv 
diagrams could be attributed to the intrinsic photometric variability 
originating from magnetic activity of one or both components.
\citep{Balaji_2015MNRAS.448..429B}. Comparing etv diagrams of KIC\,4832197 
and a very active eclipsing binary KIC\,12418816 \citep[][see Fig. 1 
in their study]{kic12418816_2018MNRAS.474..326D}, it is clearly seen 
that our target system does not exhibit clear patterns, which indicates 
much lower level of magnetic activity.

Combining the best-fitting spectroscopic orbit and light curve 
models, absolute physical properties of the components of the 
system are computed. Using these parameters, components are 
plotted on $Log~T_{eff}-Log~L/L_{\odot}$ plane. Positions of the 
components on this plane suggests 2.8$\pm$0.8 Gyr age for the 
system. The primary component of KIC\,4832197 appears to burnt
almost half of its main sequence fuel.

$UBV$ colours \citep{ubv_kepler_2012PASP..124..316E} of the 
system enables to estimate interstellar reddening. Trial 
plotting of the $U-B$ and $B-V$ colours of the system on
$UBV$ colour-colour diagram yields $E(B-V)=0\fm009$. This
indicates 452$\pm$40 pc of distance. Considering the reported 
observational errors of $UBV$ colours and small amount of 
$E(B-V)$, interstellar reddening can be neglected. In this 
case, the distance becomes 459$\pm$40 pc. Two distance values
agrees within the computed statistical error. Computed error 
of the distance is mainly dominated by the estimated 200 K 
uncertainty of the effective temperatures of the components.
The distance of the system based on precise parallax measurement
of GAIA \citep{GAIA_first_paper_2016A&A...595A...1G, 
GAIA_DR3_2022arXiv220800211G, 
GAIA_DR3_Catalogue_Validation_2022arXiv220605989B} is given
as 449$\pm$2 pc, which is in good agreement with the computed
distance in this study.

Removing the best-fitting light curve model from long cadence
Kepler light curve, residual light curve is obtained, which 
shows clear brightness variability at out-of-eclipse phases. 
Residual light curve exhibits remarkable changes, which has a 
time scale of a few orbital cycle. These changes occur both in 
amplitude and shape of the light curves. For instance, an 
asymmetric single-humped light curve can become a double-humped
one with a noticeable amplitude change in a few days and then
can restore itself to an another asymmetric light curve with
a different amplitude. Considering spectral types of the components 
the most likely explanation for the source of the out-of-eclipse 
brightness variability is cool spot activity on one or both 
components. Such a spot activity is the result of magnetic 
activity in cool stars and causes to appear emission features 
in particular activity-sensitive spectral lines through the 
optical spectrum (e.g. H$\alpha$, Ca\,{\sc ii} H\& K lines). 
However, no emission features are observed neither in H$\alpha$ 
nor in Ca\,{\sc ii} H\& K lines in TFOSC spectra of KIC\,4832197. 
Nevertheless, several very weak stellar flares, which can be 
considered photometric evidence of cool spot activity, are 
detected in long cadence Kepler light curve of the system. Such a 
situation was observed before for eclipsing binary KIC\,9451096
\citep{KIC9451096_2018RMxAA..54...37O}, where no emission was 
detected in activity sensitive lines but significant number of 
flares were detected in Kepler light curve. On the other hand,
although the amplitude of the rotational modulation signal reaches 
up to 0.01 in normalized flux unit, this is still a low amplitude 
brightness variability, thus observing no emission in 
activity-sensitive spectral lines may not be an unexpected case. 
Actually, low level of magnetic activity is in agreement with 
physical properties of the system. Masses of the components are 
slightly higher than the Sun, which means that both of them 
possess more shallow outer convective envelope compared to the Sun. 
Therefore, one may expect decrease in the level of magnetic 
activity. Comparing KIC\,4832197 with KIC\,12418816\citep{kic12418816_2018MNRAS.474..326D}, 
which is a lower mass eclipsing binary with shorter orbital period 
and high level of magnetic activity, confirms this situation. 
Components of KIC\,12418816 possess deeper convective envelope 
compared to the components of our target. As a result of 
combination of faster rotation and deeper convective envelopes, 
high level of magnetic activity is not surprising in KIC\,12418816. 
In the case of KIC\,4832197, the situation is opposite, hence low 
level of magnetic activity is not an unexpected situation for 
KIC\,4832198.

Frequency analysis of residual light curve shows two dominant 
frequency peaks around the orbital frequency (see amplitude 
spectrum plotted in Figure~\ref{Fig_periodogram}). However, 
lowermost panel of the figure clearly shows that almost no signal 
appears at the exact location of the orbital frequency. It means 
that the best-fitting light curve model precisely represents 
binarity effects. As a result, removing the best-fitting light 
curve model from observations perfectly removes the binarity 
effects from long cadence light curve. Comparing frequencies 
($f_{1}$, $f_{2}$ and $f_{orb}$) and their statistical error 
listed in Table~\ref{Table_frequencies}, one can notice that the 
separation of frequencies exceeds the statistical errors. 
No major splitting is observed for the peaks of $f_{1}$ and $f_{2}$, 
except a remarkable side component of $f_{1}$ located at 
slightly higher frequency compared to $f_{1}$. We interpret 
that these frequencies are the results of two separate spots 
or spot groups located at different latitudes on the surface of 
one or both components. Looking at lowermost panel of the Figure~\ref{Fig_periodogram}, it is quickly seen that $f_{1}$ 
and $f_{2}$ frequencies are smaller and larger than $f_{orb}$ 
respectively. Assuming that both components possess solar type 
differential rotation on their surfaces, spot (or spot group) 
causing variability with $f_{1}$ frequency is located at lower 
latitudes, which is expected to rotate faster than the latitude 
rotating with $f_{orb}$. That latitude that rotates with $f_{orb}$
is often called co-rotation latitude). In this case, the other 
spot (or spot group), which rotates with $f_{2}$ frequency must 
be located at higher latitudes compared to the co-rotation latitude.
These latitudes are expected to rotate slower than the co-rotation 
latitude as well. Further inspection on these frequencies reveals 
that $f_{1}$ persistently appear in each separate Kepler quarter 
while $f_{2}$ weakens or disappears occasionally. In this case, 
spot or spot groups related to $f_{1}$ frequency is more persistent 
and exhibit stronger activity compared to the spot or spot group 
related to $f_{2}$ frequency. Four years of persistence of $f_{1}$ 
frequency is not surprising in the scope of solar and stellar 
connection since similar persistent active longitudes are 
reported previously for our Sun
\citep{Berdyugina_persistent_active_longitudes_2003A&A...405.1121B}.

With the current data, it is difficult to distinguish which component 
possesses spot(s) on its surface. However, we may make an implicit 
assumption by considering two mechanisms, depth of the outer convective 
envelope and the rotation, which trigger the magnetic activity on cool 
stars. Although the system is composed of two nearly similar stars, 
the secondary component is 500 K cooler than the primary. It means that 
the secondary star possesses deeper convective envelope than the primary 
star. Considering that the axial rotation period of the components are 
synchronized to the orbital period, this finding indicates that the 
secondary component may exhibit stronger magnetic activity than the 
primary component, because of its deeper convective outer envelope.
Then, we may assume that the origin of spots is the cool secondary
component.

Assuming that both $f_{1}$ and $f_{2}$ come from spot or spot 
groups on the cool secondary component, a lower limit for differential 
rotation can be set. Following the procedure described in \citet{Hall_Busby_1990_difrot}, we compute corresponding period 
for $f_{1}$ and $f_{2}$, and adopt them as minimum ($P_{min}$) and 
maximum ($P_{max}$) periods. Under solar type differential rotation 
assumption, $P_{min}$ can be adopted as equatorial rotation period. 
Then, relative surface shear can be computed via 
$P_{max} - P_{min} / P_{min}=kf$, where $k$ is the differential 
rotation coefficient and $f$ is a constant depending on the range 
of spot forming latitudes. Assuming that $f$ varies between 0.5 and 
0.7, which corresponds to 45 degrees of latitude range for spot 
forming, $k$ varies between 0.14 and 0.10, which indicates an average 
$\bar{k}=0.12$. This is half of the solar differential rotation 
coefficient ($k_{\odot} = 0.189$, see \citet{Hall_Busby_1990_difrot} 
and references therein) but twice of the differential rotation 
coefficient of the eclipsing binary KIC\,9451096 \citep[$k=0.069$,][]{KIC9451096_2018RMxAA..54...37O}. This finding 
appears contradictory to the theory of stellar magnetic activity 
since KIC\,9451096 possesses higher level of magnetic activity but 
weaker differential rotation compared to our target system in this 
study.

Increasing the number of precisely analysed solar type 
eclipsing binaries will provide a global look to the photometric
properties of magnetic activity on eclipsing binaries and reflection
of differential rotation, which is one of the two key mechanisms 
responsible for magnetic activity in cool stars, on light curves.
Continuous photometric data provided by space telescopes is the 
key to achieve this purpose.

\section*{Acknowledgments} 
We acknowledge anonymous referee for his/her valuable comments that 
improved the quality of the paper. We thank T\"UB\.ITAK for partial support 
in using RTT150 (Russian-Turkish 1.5-m telescope in Antalya) with project 
number 14BRTT150-667. This paper includes data collected by the Kepler 
mission. Funding for the Kepler mission is provided by the NASA Science 
Mission Directorate.

\bibliography{kic4832197}

    

\end{document}